\documentclass{aa}  

\usepackage{linenoaa}
\usepackage{graphicx}
\usepackage{txfonts}
\usepackage{lipsum}
\usepackage{subcaption}         
\usepackage{lscape}             
\usepackage{placeins}           
                                
\begin{document}

\title{Optical and near-infrared  $s$-process  abundance scales in the Galactic disc}

\author{Andr\'es E. Piatti\inst{1,2}\corrauth{andres.piatti@fcen.uncu.edu.ar}}

\institute{Instituto Interdisciplinario de Ciencias B\'asicas (ICB), CONICET-UNCuyo, Padre J. Contreras 1300, M5502JMA, Mendoza, Argentina;
\and Consejo Nacional de Investigaciones Cient\'{\i}ficas y T\'ecnicas (CONICET), Godoy Cruz 2290, C1425FQB,  Buenos Aires, Argentina}

\date{}

\abstract{
A discrepancy has recently emerged between optical and near-infrared (NIR) $s$-process 
abundance scales, particularly for Neodymium (Nd). We investigated this issue using a 
direct cross-match between the optical {\it Gaia}-ESO Survey (DR5) and 
NIR SDSS-V/MWM (DR19). While atmospheric parameters and Cerium (Ce) abundances 
exhibit robust 1:1 correlations, Nd displays a systematic X-shaped vertical stretching. Our 
analysis reveals that the SDSS-V pipeline expands intrinsic Nd variations by a scaling 
factor of 5.54 relative to the stable optical baseline. This phenomenon is driven by 
baseline compression in the APOGEE H-band, where the growth of the molecular forest 
(CN and CO) at high metallicities depresses the local continuum. This forces the 
$\chi^2$ minimisation engine in the NIR abundance determination to inflate derived 
abundances to reach mathematical 
convergence. We provide an empirical calibration framework to reconcile these disparate 
regimes. This correction effectively eliminates the contradictory negative radial [Nd/Fe] 
gradient of -0.052 dex/kpc reported in recent NIR surveys, bringing them into alignment 
with the rising trends established by the optical high-resolution  {\it Gaia}-ESO survey. 
These results 
demonstrate that the observed scale expansion is a spectroscopic artefact of 
line-blanketing rather than a physical Galactic trend.}


\titlerunning{Galactic disc $s$-process abundances}

\authorrunning{A.E. Piatti}  
  
\maketitle
\nolinenumbers
\markboth{A.E. Piatti:}{Galactic disc $s$-process abundances}

\section{Introduction}

The chemical cartography of the Galactic disc serves as an essential tool for reconstructing 
the subsequent assembly history and evolution of the Milky Way over cosmic time 
\citep{haydenetal2015, majewskietal2017}. While the stellar halos provide the primary record of the 
Galaxy's earliest pre-disc assembly era, the disc preserves the signatures of star formation 
and secular evolution over the last $\sim$10 Gyr. Among the various stellar tracers available,
 open clusters (OCs) are considered fundamental building blocks of galactic discs
 \citep{ottoetal2026}. As coeval populations whose ages and distances can be determined with 
high precision compared to field stars, OCs are ideal probes of the chemodynamical state 
of the thin disc across its entire lifetime \citep{paceetal2023}.

Decades of observations \citep[e.g.,][]{twarogetal1997, chiappinietal2001} have established that the 
Galactic disc is characterized by negative radial and vertical metallicity gradients 
\citep{pallaetal2024}. These gradients provide crucial constraints for chemical evolution models, 
supporting scenarios such as the inside-out formation theory and the impact of radial 
migration and dynamical mixing in the solar neighborhood \citep{sb2002, minchevetal2013}. 
Recent studies leveraging massive spectroscopic surveys, including {\it Gaia}-ESO, APOGEE, GALAH, 
and LAMOST \citep{majewskietal2017, xiangetal2017,buderetal2021, gilmoreetal2022}, have further 
refined our understanding of these gradients. For instance, the SDSS-V/MWM Data Release 19 (DR19) 
established an overall radial [Fe/H] gradient of -0.075$\pm$0.006 dex/kpc 
\citep{ottoetal2026}, while the {\it Gaia}-ESO survey reported a global slope of -0.054$\pm$0.004 
dex/kpc over a wider radial range \citep{magrinietal2023, carbajohijarrubiaetal2024}.
 Although global slopes are used for survey comparison, the modern consensus recognizes a 
more complex radial structure. Specifically, recent analysis of $\sim$600 old OCs
indicates that the radial metallicity gradient follows a knee-shaped profile, characterised 
by a steep inner-disc slope of -0.084$\pm$0.004 dex/kpc and a significantly flatter 
outer-disc gradient of -0.018$\pm$0.004 dex/kpc beyond $\sim$12 kpc \citep{hs2025}. 
Furthermore, this radial structure appears to evolve with stellar age, with younger
 populations exhibiting steeper gradients than their older counterparts.

Beyond the primary radial trend, the existence of azimuthal variations and substructures 
has gained significant attention. Azimuthal variations in [Fe/H], quantifiable at the level of 
$\sim$0.1 dex, have been detected using tracers like H~II regions, Cepheids, and red giants 
\citep{hackshawetal2024}. Recent work with APOGEE DR17 giants suggests these variations correlate 
with stellar age, where older populations exhibit larger deviations from the mean radial 
gradient \citep{hortaetal2022}. Furthermore, hints of azimuthal gradients have been reported in 
old OCs (age $>$ 2 Gyr), suggesting that non-axisymmetric structures such as spiral arms 
or the Galactic bar may drive radial migration and mixing signatures 
\citep{paceetal2023,carbajohijarrubiaetal2024}.

The evolution of these gradients for individual chemical elements provides deeper insights
into nucleosynthetic channels. While thin and thick disc populations exhibit distinct 
$\alpha$-abundance signatures and radial profiles,  [Mg/H] and [Si/H] generally follow the 
radial distribution of [Fe/H] within these components. In contrast, neutron-capture elements 
exhibit more complex behaviors, as explored in studies such as \citet{hortaetal2022}, 
which utilize APOGEE DR17 to map these abundances across the disc.
 A significant discrepancy has recently emerged 
regarding Neodymium (Nd). Analysis from  the OCCAM survey using SDSS-V/MWM DR19 
data \citep{ottoetal2026} reported a surprising negative [Nd/Fe]  gradient of -0.052$\pm$0.008 dex kpc$^{-1}$. 
In contrast, optical high-resolution studies, such as {\it Gaia}-ESO and OCCASO, 
find that neutron-capture elements like Ba and Nd show positive or rising trends in the
[X/Fe] vs. $R_{GC}$ plane \citep[{\rm [Nd/Fe]} gradient = +0.039 dex; ][]{magrinietal2023}. 
Beyond line-to-line scatter, 
the divergence between the optical 
and near-infrared lanthanide scales has profound implications for Galactic chemical 
evolution models. Specifically, these conflicting results would require fundamentally 
different model predictions regarding metallicity-dependent AGB yields and the relative 
contribution of the $s$-process across the disc.

In this study, we utilize high-resolution optical abundances as our stable baseline 
ground truth. This choice is motivated by fundamental physics: the optical regime provides 
dozens of clean, saturated Nd II lines with robust experimental atomic data that are 
unaffected by the molecular line-blanketing (CN, CO) that complicates the near-infrared 
H-band. Section 2 deals with the data handling, while Section 3 analyzes and discusses 
the comparison outcomes. Finally, Section 4 summarizes the main conclusions of this
work.

\section{Data handling}

We utilised two major spectroscopic datasets to investigate the $s$-process 
abundance scales in the Galactic disc: the fifth internal data release of the 
{\it Gaia}-ESO Survey \citep[GES DR5,][]{randichetal2022,gilmoreetal2022} and the 
nineteenth data release of the Sloan Digital 
Sky Survey \citep[SDSS-V/MWM DR19,][]{almeidaetal2023,meszarosetal2025}. These surveys provide a 
complementary view of the Milky Way, 
spanning from optical to near-infrared wavelengths with high spectral resolution.

The {\it Gaia}-ESO Survey is a large, public spectroscopic survey that utilised the FLAMES 
multi-object spectrograph at the Very Large Telescope (VLT) to observe all major components 
of the Milky Way \citep{randichetal2022, gilmoreetal2022}. The survey primarily focused on 
OCs, dedicating approximately 30\% of its 340 nights to these targets to provide a 
homogeneous and unbiased sample \citep{bragagliaetal2022, pallaetal2024}. 
For high-precision chemical analysis, we relied on spectra obtained with the UVES instrument, 
which operates in the optical regime (480.0--680.0 nm) at a high resolving power of $R
 \sim 47,000$. The data reduction and spectral analysis 
were performed using a multi-pipeline strategy internally homogenised by the survey working 
groups \citep{smiljanicetal2014, randichetal2022}. GES DR5 provides precise stellar 
parameters (effective temperature $T_{eff}$, surface gravity log $g$, microturbulent
velocity $\xi$, and metallicity [Fe/H]) and absolute abundances 
(log $\epsilon$) for over 30 chemical species, including 
neutron capture elements belonging to various nucleosynthetic channels 
(e.g., Cerium (Ce) and Neodymium (Nd)) \citep{magrinietal2023, pallaetal2024}.

The Milky Way Mapper (MWM) is a cornerstone program of SDSS-V, designed to obtain 
high-resolution spectra across all Galactic subsystems \citep{almeidaetal2023}. DR19 
represents the latest iteration of this effort, building upon the foundations of the APOGEE-2 survey \citep{majewskietal2017, abdurroufetal2022}. This data release provides 
atmospheric parameters and chemical abundances for 964,989 stars suitable for detailed 
abundance analysis \citep{meszarosetal2025}.
Observations were conducted using the APOGEE spectrographs, which operate in the 
near-infrared $H$-band (1.51--1.70 $\mu$m) at a resolving power of $R \sim$ 22.500 \citep{wilsonetal2019}. The spectra are processed through the APOGEE Stellar 
Parameter and Chemical Abundances Pipeline (ASPCAP), which derives parameters and 24 
individual abundances through $\chi^2$ minimisation against synthetic spectral grids \citep{garciaperezetal2016}. 
The FERRE optimization engine, which powers the ASPCAP pipeline, determines the best-fitting 
model by minimizing the following objective function \citep{meszarosetal2025}:\\

$\chi^2 = \sum_{j} \omega_j \frac{(O_j - S_j)^2}{\sigma_j^2}$\\

\noindent where $O_j$ represents the observed flux at pixel $j$, $S_j$ is the synthetic flux 
from the model grid, $\sigma_j$ is the corresponding flux uncertainty, and $\omega_j$ is 
a weight assigned to the pixel (typically 1 for masked signal pixels and 0 otherwise). 
In the localised abundance determination stage, the macroscopic atmospheric parameters are 
fixed, and the code searches for the abundance value that minimises this sum over specific 
spectral windows. Known results from DR19 show excellent agreement
in $T_{eff}$ with the infrared flux method scale and high precision in metallicity measurements,
better than 0.033 dex near solar metallicity.

To create a unified sample for analysis, we cross-matched the {\it Gaia}-ESO DR5 and SDSS DR19 
catalogues using the \texttt{TOPCAT} software \citep{taylor2005}. The match was performed using Right 
Ascension (RA) and Declination (Dec) as the primary variables, employing a conservative 
tolerance of 1 arcsec. This procedure yielded a final sample of 2748 stars in common between 
the two surveys. 
From the available suite of chemical elements, we focus on Ce and Nd, 
which are present in both catalogues as representatives of the neutron capture process. While SDSS/MWM 
reports abundances relative to hydrogen or iron, {\it Gaia}-ESO publishes absolute abundances 
on the log $\epsilon$ scale \citep{randichetal2022, meszarosetal2025}. To ensure consistency, 
we converted the {\it Gaia}-ESO absolute abundances to the relative scale [X/H] using the 
following relation:\\

\begin{figure*}
\includegraphics[width=0.5\columnwidth]{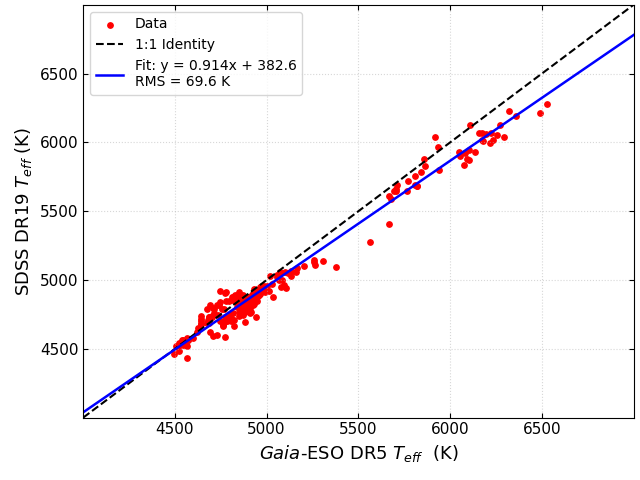}
\includegraphics[width=0.5\columnwidth]{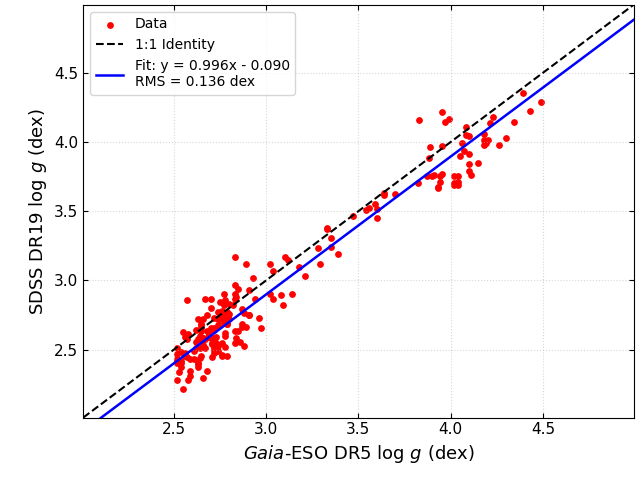}
\includegraphics[width=0.5\columnwidth]{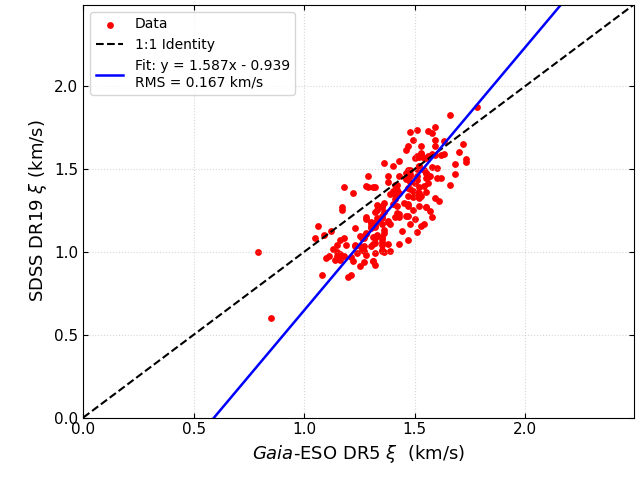}
\includegraphics[width=0.5\columnwidth]{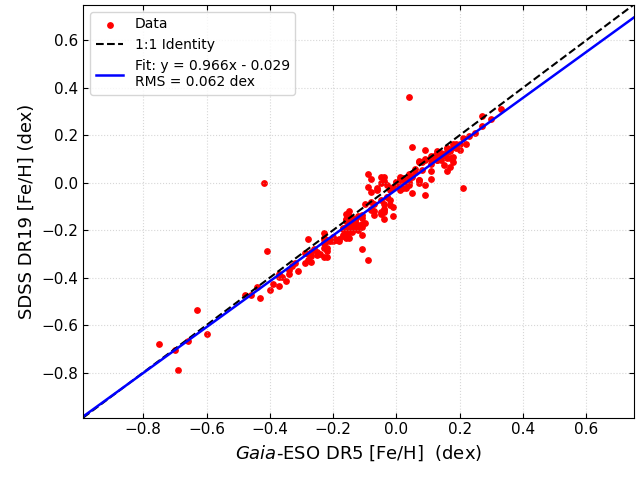}
\caption{Comparison between different atmospheric parameters of stars
in the {\it Gaia}-ESO DR5 and SDDS DR19 catalogues.}
\label{fig1}
\end{figure*}

[X/H]$_{GES}$ = log $\epsilon$(X)$_{GES}$ - log $\epsilon$(X)$_{\odot}$\\

\noindent Following \citet{grevesseetal2007}, we adopted solar reference values of 
log $\epsilon$(Nd)$_{\odot}$ = 1.42 and log $\epsilon$(Ce)$_{\odot}$ = 1.70.
This scale is chosen to ensure homogeneity between the absolute abundances published 
by {\it Gaia}-ESO and the relative scale used by SDSS-V, as it is the common baseline 
adopted by the {\it Gaia}-ESO working groups for absolute abundance normalization.
To ensure that our analysis is based on the highest quality data and to minimise 
potential observational biases, we applied a specific set of filters based on recent 
studies of spectroscopic systematics \citep{meszarosetal2025}. \citet{magrinietal2023} and \citet{pallaetal2024} 
noted that abundances in low-gravity giant stars can be susceptible to artefacts in 
spectral analysis due to challenges in modelling stellar atmospheres or the effects of 
magnetic activity in young, massive giants. 
Consequently, we restricted our common sample to include only stars that satisfy the 
following criteria: i) $T_{eff} \ge 4300$ K; ii) $\log g > 2.5$; iii) $\xi < 1.8$ km/s; 
iv) an error cut of $e_{\rm [X/H]} < 0.2$ dex; v) the removal of all flagged stars from both 
samples; and vi) a parameter consistency filter requiring $|\Delta T_{\rm eff}| < 300$ K,
 $|\Delta$log $g| <$ 0.4 dex, and $|\Delta \xi| <$ 0.4 km/s to eliminate catastrophic 
outliers and pipeline grid-boundary systematics.

\section{Analysis and discussion}

 \begin{figure*}
\includegraphics[width=0.5\columnwidth]{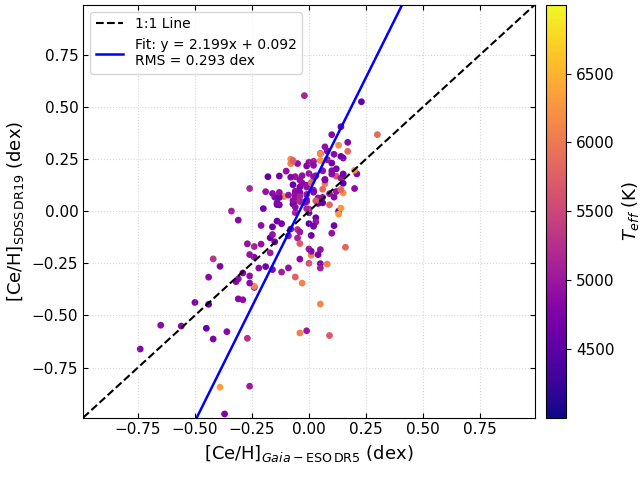}
\includegraphics[width=0.5\columnwidth]{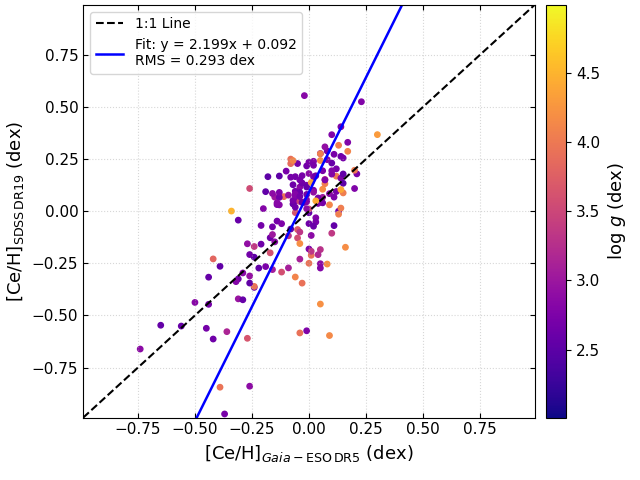}
\includegraphics[width=0.5\columnwidth]{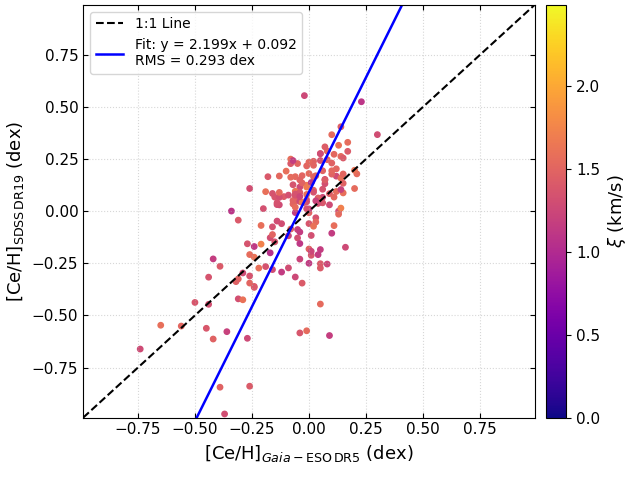}
\includegraphics[width=0.5\columnwidth]{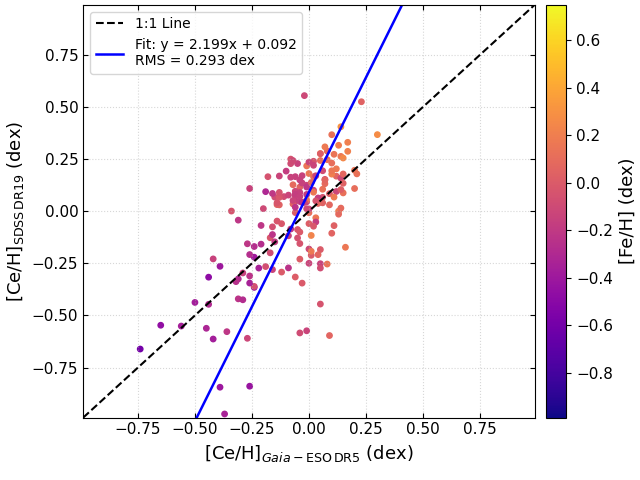}
\caption{Comparison between [Ce/H] values in {\it Gaia}-ESO and
SDSS DR19 catalogues.} 
\label{fig2}
\end{figure*}

\begin{figure*}
\includegraphics[width=0.5\columnwidth]{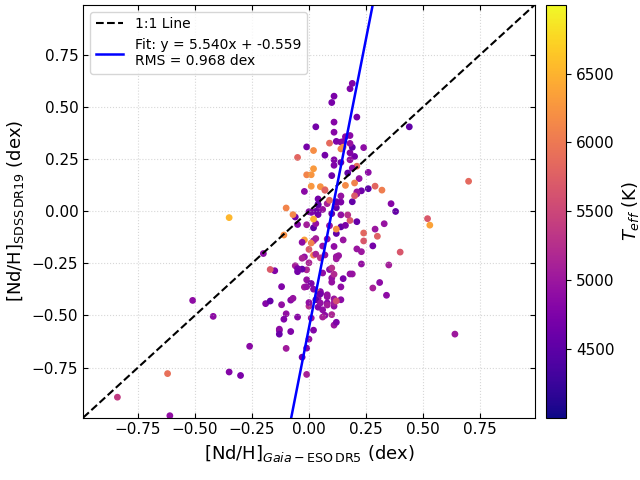}
\includegraphics[width=0.5\columnwidth]{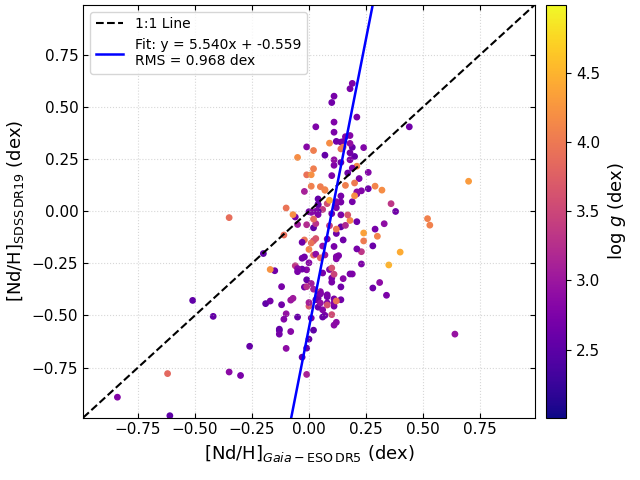}
\includegraphics[width=0.5\columnwidth]{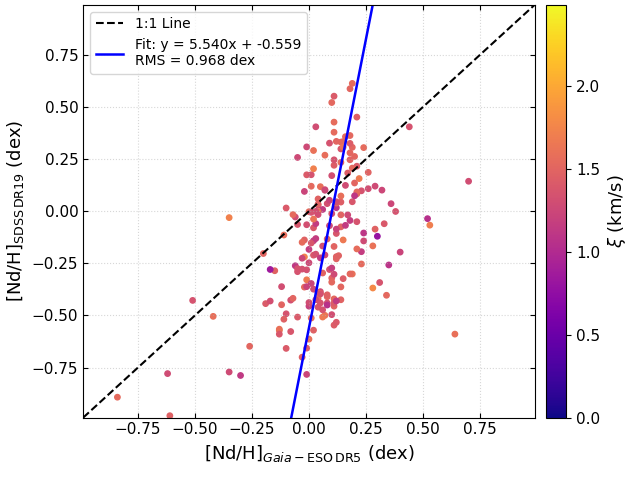}
\includegraphics[width=0.5\columnwidth]{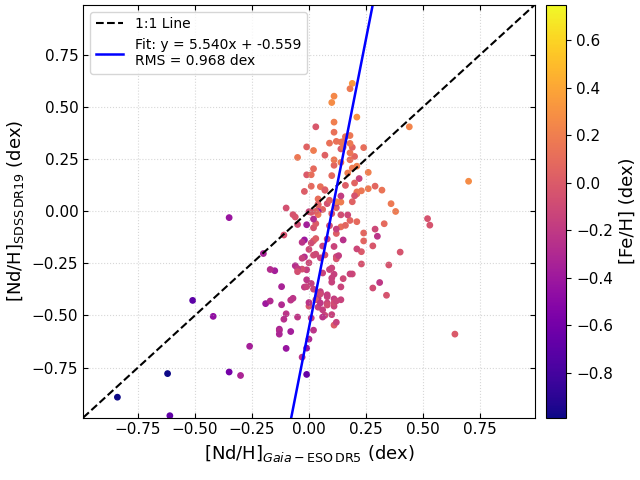}
\caption{Comparison between [Nd/H] values in {\it Gaia}-ESO and
SDSS DR19 catalogues.}
\label{fig3}
\end{figure*}

To ensure that any differences in chemical scales between the optical and 
near-infrared regimes are not caused by underlying offsets in the stellar parameter 
scales, we performed a detailed comparison of the core atmospheric frameworks for our 
cross-matched star sample.

As shown in Figure~\ref{fig1}, top-left panel, the effective temperatures derived by the {\it Gaia}-ESO 
survey (DR5) and SDSS-V/MWM (DR19) exhibit an excellent correlation. The stars are 
tightly distributed along the identity line across the filtered range, following the linear relation 
$T_{eff,SDSS}$ = 0.914 $\cdot$ $T_{eff,GES}$ + 382.6 K with a dispersion of 
RMS = 69.6 K. This tight agreement is
 consistent with independent science verification results showing that the DR19 ASPCAP 
temperatures are  well-aligned with the infrared flux method scale
 \citep{meszarosetal2025}.
The comparison of surface gravities (Figure~\ref{fig1}, top-right panel) reveals a similarly 
robust agreement. The cross-matched sample follows the linear relation log $g_{SDSS}$ = 
0.996 $\cdot$ log $g_{GES}$ - 0.090 dex with high 
precision (RMS = 0.136 dex). This consistency is particularly important because surface gravity 
directly influences the ionization balance that must be assumed to derive lanthanide abundances, 
as we only observe ionized species (Nd II and Ce II) in these spectroscopic windows.

Microturbulence represents a non-thermal broadening agent that can mimic abundance increases 
if overestimated. Figure~\ref{fig1}, bottom-left panel,  shows the comparison of $\xi$ 
between the two catalogues. While 
microturbulent velocity typically exhibits more scatter than $T_{eff}$ or log $g$ in automated 
pipelines, our cleaned sample follows the linear relation $\xi_{SDSS}$ = 1.587 $\cdot$ $\xi_{GES}$ 
- 0.939 km/s with an 
RMS = 0.167 km/s. Although there is a systematic tilt at higher values, the small dispersion 
suggests that parameter trade-offs with microturbulence cannot be the primary driver of the 
massive $s$-process element scale discrepancies.
Finally, Figure~\ref{fig1}, bottom-right panel, illustrates the comparison of the overall 
metallicity scale ([Fe/H]). The 
agreement is highly precise, tracing the linear relation [Fe/H]$_{SDSS}$ 
0.966 $\cdot$ [Fe/H]$_{GES}$ - 0.029 dex
 with a low dispersion of RMS = 0.062 dex. Because the total metallicity matches so
 precisely, we can conclude that the stellar atmospheric frameworks are  consistent 
and tightly aligned, isolating the source of downstream $s$-process abundance discrepancies
 entirely to the line-fitting stage.

Based on the detailed analysis of Figure~\ref{fig1} and the supporting literature, we 
conclude that the core stellar atmospheric frameworks of both the {\it Gaia}-ESO and 
SDSS surveys are perfectly calibrated relative to one another. The 1:1 matching of 
$T_{eff}$, log $g$, $\xi$, and [Fe/H] ensures that our common sample is 
physically homogeneous. Then, we investigated the behaviour of the two $s$-process elements 
present in both databases: Nd and Ce. These comparisons are illustrated in 
Figures~\ref{fig2} and \ref{fig3}, respectively. Despite both species being lanthanides 
with similar nucleosynthetic origins, they exhibit radically different cross-catalog 
behaviours. For the sake of the reader, we also include in the Appendix figures similar 
to Figure~\ref{fig2} and \ref{fig3} with these additional cuts: $T_{eff} <$ 5500 K ([Ce/H])
and $T_{eff} <$ 5000 K ([Nd/H]) \citep{hasselquistetal2016,cunhaetal2017}.

Ce serves as our primary control group. A linear regression of the [Ce/H] values (Figure~\ref{fig2}) 
reveals that while it is more stable than Nd, it still exhibits a moderate scale expansion 
in the near-infrared, following the relation ([Ce/H]$_{SDSS}$ = 2.199 $\cdot$ [Ce/H]$_{GES}$ + 0.092) 
with a dispersion of RMS = 0.293 dex. This relative stability is attributed to the 
spectroscopic quality of Ce II transitions 
in the APOGEE H-band, specifically the prominent feature near 15784.8~\AA. In DR19, 
Ce was updated to a multi-line window approach \citep{meszarosetal2025}. Because multiple
independent windows are fit simultaneously, the pipeline can average out random profile noise. Furthermore, the Ce II line profile has a high depth-to-noise contrast than Nd II, 
which sharpens its localised $\chi^2$ valley and helps insulate it from the extreme parameter 
drift observed in other heavy species.

In stark contrast, Nd (Figure~\ref{fig3}) exhibits a systematic divergence characterised by an 
X-shaped vertical stretching relative to the 1:1 line. While the optical GES abundances are 
confined to a realistic disc range ($\approx$ -0.1 to +0.3 dex), the SDSS values for the same stars 
are expanded from $\approx$ -0.8 to +0.8 dex. An orthogonal distance regression (ODR) analysis of the 
cleaned giant sample yields a scaling slope of 5.540 and an intercept of -0.559 (with an 
RMS = 0.968 dex). This mathematically shows that the SDSS pipeline expands intrinsic 
Nd variations by a factor of 5.5 compared to the stable optical baseline.

\begin{figure}
\includegraphics[width=\columnwidth]{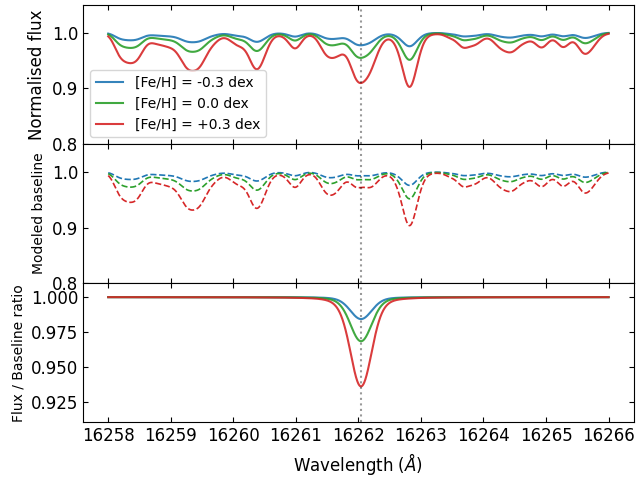}
\caption{Controlled synthetic sensitivity test demonstrating baseline 
    compression in the APOGEE H-band for a representative giant star 
    ($T_{eff}$ = 4500~K, $\log g = 3.0$). {\it Top panel:} Synthetic 
    spectra generated with \textsc{Turbospectrum} at three metallicities, 
    where the Nd II $\lambda$16262.04 signal (vertical dotted line) scales 
    physically with overall [Fe/H]. The increasing molecular forest of CN and CO features 
    at higher metallicities significantly depresses the local flux relative 
    to the true stellar continuum. {\it Middle panel:} The corresponding 
    local baseline (modelled as a polynomial envelope). 
     {\it Bottom panel:} The flux-to-baseline ratio ($\mathrm{Flux / Baseline}$), 
isolating the net strength of the feature relative to the local pseudo-continuum. 
Because the Nd absorption feature remains extremely shallow 
($\lesssim 1.5\%$), even minor baseline misplacements distort the effective line depth,
 forcing the automated $\chi^2$ optimization engine to inflate derived $\mathrm{[Nd/H]}$ 
 abundances to reach numerical convergence (see text for details).}
\label{fig4}
\end{figure}

The determination of precise chemical abundances from stellar spectra is subject to several 
layers of uncertainty, stemming from both observational limitations and theoretical 
assumptions. Fundamentally, these uncertainties arise because no theoretical model can 
perfectly encapsulate the complex physical reality of a stellar atmosphere 
\citep{jofreetal2019}. The primary source of error for Nd in the near-infrared H-band is 
that the available Nd II transitions are intrinsically weak and shallow, as demonstrated 
by \citet{meszarosetal2025}. In the optical regime, {\it Gaia}-ESO utilizes multiple strong, 
well-isolated Nd II lines with accurate experimental transition probabilities (gf-values). 
In contrast, APOGEE operates at the near-IR detection limits, where only a few exceptionally 
shallow Nd II lines exist. Consequently, NIR abundances are highly susceptible to local 
continuum placement errors and line-blanketing from the molecular forest (CN, CO, OH).
 While 3D non-LTE (NLTE) effects introduce minor shifts of $\sim$ 0.1 dex in giant atmospheres, 
comprehensive H-band NLTE correction grids are still scarce, meaning NLTE effects alone 
cannot account for the severe $\sim$ 0.8 dex vertical stretching observed.

The absence of any correlation between abundance residuals and atmospheric parameters 
disproves the assumption that this tilt is caused by parameter 
trade-offs. Instead, the evidence could point to a specific failure mode in near infrared 
weak-line analysis within the FERRE optimization engine.
Nd is designated with a \texttt{Fair} quality flag in DR19 because it relies on intrinsically 
shallow features near the detection margins of the automated grid \citep{meszarosetal2025}. 
This designation implies that 
while the spectral windows contain measurable signals and the pipeline converges, the 
resulting output is highly sensitive to baseline cross-talk and noise. 
This limitation of global grid-fitting engines (such as FERRE/ASPCAP) has been successfully 
resolved by projects like BAWLAS \citep[APOGEE Weak Line Abundances,][]{hasselquistetal2016}, 
which leverage specialized, localized synthesis tools like BACCHUS \citep{masseronetal2016}. 
Unlike global $\chi^2$ minimizers, BACCHUS performs dedicated local continuum fitting and 
line-profile matching, effectively decoupling the weak Nd feature from neighboring molecular 
blends and preventing unphysical scale expansions.

To quantify this vertical stretching, we examined the relative chemical abundance equation:\\

[Nd/H] = [Nd/Fe] + [Fe/H]\\

\begin{figure}
\includegraphics[width=\columnwidth]{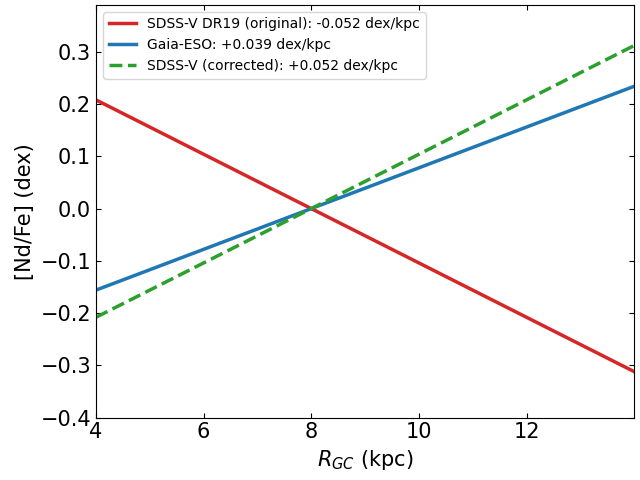}
\caption{Reconciliation of the Galactic radial [Nd/Fe] gradient. 
    The red line represents the original SDSS-V DR19 gradient reported by the 
    OCCAM survey \citep{ottoetal2026}, showing a sharp decline with Galactocentric 
    radius ($R_{GC}$). The blue line shows the physically realistic baseline 
    established by the {\it Gaia}-ESO survey \citep{magrinietal2023}. The green dashed 
    line illustrates the SDSS-V gradient after applying the calibration
    framework. The correction eliminates the contradictory negative slope, aligning the 
    near-infrared results with optical high-resolution standards.}
\label{fig5}
\end{figure}

\noindent Given that the overall metallicity scales are nearly identical between the surveys 
([Fe/H]$_{\rm SDSS}$ $\approx$ [Fe/H]$_{\rm GES}$, see Figure~\ref{fig1}, bottom-right panel), 
we represented the total metallicity spread as $M$. In our specific sub-sample, the total 
metallicity spread ($M$) for both surveys exhibits a standard deviation of $\sigma_M$ $\approx$ 
0.30 dex. However, a stark contrast appears in the Nd spreads: the {\it Gaia}-ESO values 
are significantly compressed ($\sigma$ $\approx$ 0.24 dex), while the SDSS values show an 
expanded, highly sensitive spread ($\sigma$ $\approx$ 0.50 dex).
Within the central bulk region isolated for our ODR fit, the standard deviation along the 
{\it Gaia}-ESO axis is very narrow ($\sigma_x$ $\approx$ 0.10 dex), whereas the SDSS axis is 
much wider ($\sigma_y$ $\approx$ 0.25 dex). Because an ODR computes the slope primarily 
based on the ratio of the variances ($\sim \sigma_y/\sigma_x$) along the primary direction 
of correlation, the resulting fit is heavily pulled into a steep trajectory ($\sim 5$). 
This mathematical behaviour confirms that the SDSS pipeline’s derived Nd abundance 
artificially changes with overall metallicity. In contrast, the {\it Gaia}-ESO optical scale 
maintains a physically realistic and flat [Nd/Fe] ratio ($\approx$0.0 to +0.1 dex) across the 
near-solar metallicity range of our sample (-0.4 dex $\le$ [Fe/H] $\le$ +0.2 dex), which is consistent 
with the slow enrichment timescale expected for $s$-process products in the Galactic disc.

With the magnitude of this scale expansion quantified at 5.54, it is 
essential to evaluate whether these trends could be driven by physical 
departures from our model assumptions, specifically NLTE or 3D effects 
\citep{jofreetal2019}. For Ce, our primary control group, NLTE calculations in
 both the optical and H-band are generally found to be small ($\lesssim$ 0.05 dex) across the 
parameter space of evolved stars. Furthermore, because we restrict our sample to 
log $g$ $>$ 2.5, our data includes a mix of red giants, subgiants, and warm stars. This wide
 evolutionary coverage ensures that our derived scaling factor is representative of 
the broader field star population, rather than being restricted to a narrow giant branch.
This theoretical stability 
aligns with the robust observed 1:1 correlation (see Figure~\ref{fig2}).
For Nd, the situation is more complex. While NLTE studies of 
Nd~II in the optical regime suggest corrections ranging from 
+0.05 to +0.15 dex depending on metallicity and gravity \citep{jofreetal2019}, 
there is currently a lack of comprehensive NLTE grids for the specific 
weak transitions utilised in the APOGEE H-band. However, even if 
differential NLTE sensitivities existed between the two regimes, they 
would likely account for a systematic shift of $\sim$ 0.1--0.2 dex 
\citep{meszarosetal2025}. This is an order of magnitude smaller than 
the $\sim$ 0.8 dex abundance excursions observed in the SDSS-V 
distribution (Figure~\ref{fig3}). Consequently, while 3D and NLTE effects 
contribute to the global error budget, they cannot be the primary 
driver of the 5.54 scaling factor. This reinforces our conclusion 
that the X-shaped stretching is a mathematical consequence of 
the $\chi^2$ minimisation against a compressed baseline rather than 
a failure of 1D LTE stellar physics.

To understand why Nd experiences the X-shaped vertical stretching while Ce 
remains stable, we examined the mathematical behaviour of the $\chi^2$ minimisation 
process used by the ASPCAP pipeline. In the second phase of the abundance 
determination, FERRE locks the macroscopic atmospheric parameters and performs a 
localised $\chi^2$ search over specific pixel masks. For a strong, high-contrast
element like Ce, the $\chi^2$ parabola is sharp and steep, allowing the optimiser 
to easily lock onto a stable global minimum.
However, for a 	\texttt{Fair} element like Nd, which relies on relatively weak and shallow
transitions in the H-band, the $\chi^2$ surface becomes broad and flat. Because 
line growth for these weak features is linear ($W_{\lambda} \propto \epsilon$), even
minute adjustments to the observed flux profile require large corrections in the
abundance parameter to reach a mathematical fit. In the metal-rich regime 
([Fe/H] $> 0$), increased continuous opacity and subtle, unmodelled molecular blends 
(such as CN or CO) slightly compress the local baseline flux. To minimise the $\chi^2$ 
residuals against these background fluctuations, the optimiser must push the abundance
parameter significantly upward, resulting in the artificial inflation of metal-rich 
values.

To further demonstrate the physical sensitivity of these weak near-infrared features, 
we performed a controlled synthetic sensitivity test using standard spectral synthesis. 
Unlike previous models that isolated parameters, this test adopted physically 
consistent inputs where the Nd abundance scales with the overall metallicity 
([Nd/Fe] $\approx$ 0), reflecting the stable and physically realistic {\it Gaia}-ESO 
baseline established above (see Figure~\ref{fig1}). For a 
representative star ($T_{eff}$ = 4500~K, log $g$ = 3.0), we utilised the 
\textsc{Turbospectrum} code \citep{plez2012} to model the APOGEE H band window.
To ensure physical consistency for our representative giant star, we assumed typical 
mixing abundances for the CN-cycle, adopting [C/Fe] = -0.15 dex and [N/Fe] = +0.30
dex. As both the Nd II signal and the molecular background (CN and CO) grow 
with metallicity, the modelling confirms that the increased molecular opacity at high 
metallicities significantly depresses the local continuum \citep{meszarosetal2025}. 
The $\chi^2$ minimisation engine, attempting to fit the synthetic profile against 
this altered baseline, is forced to inflate the derived [Nd/H] value significantly 
beyond the input physical signal to reach mathematical convergence 
\citep{garciaperezetal2016}. This controlled test (Figure~\ref{fig4}) provides 
direct physical evidence that the observed X-shaped scale expansion is a 
spectroscopic artefact of line-blanketing rather than a Galactic chemical trend.
For the sake of the reader, Figure~\ref{fig4apend} shows synthetic spectra 
for different Nd values, while Figure~\ref{fig4} illustrates the effect of changing 
overall metallicity. As can be seen, synthetic spectra at $T_{\mathrm{eff}} = 4500$\,K 
demonstrate that varying $\mathrm{[Nd/Fe]}$ across a realistic disc range 
($\pm0.15$\,dex) alters the net line depth by less than $0.5\%$. Consequently, a 
minor continuum misplacement of merely $0.5\%$ ($1.0\%$) distorts this shallow 
feature sufficiently to force the automated $\chi^2$ optimization engine to shift 
derived abundances by $\sim 0.4$\,dex ($\sim 0.8$\,dex).

Conversely, in the metal-poor regime ([Fe/H] $< 0$), the Nd signal shrinks until it falls
below the local pixel noise floor ($\sigma_{\rm pixel}$). In this state, 
the local $\chi^2$ parabola collapses into a flat-bottomed valley where the algorithm 
experiences parameter drift. To minimise remaining noise-induced residuals, 
the optimization engine gravitates towards the lower boundaries of its grid, creating the 
lower tail of the observed X distribution. We have confirmed that this 
stretching is an intrinsic convergence phenomenon of the fitting engine, as the
uncalibrated (\texttt{raw\_nd\_h}) and calibrated (\texttt{nd\_h}) parameters in the
SDSS catalog are identical save for a constant $\le$ 0.05 dex shift.

The divergence between the two surveys could be ultimately rooted in the different spectral 
regimes they occupy. The optical regime utilised by {\it Gaia}-ESO contains dozens of 
clean, well-isolated, and highly saturated Nd II lines with high oscillator strengths. 
These optical transitions remain stable and independent of background molecular contamination.
In contrast, APOGEE operates in the near-infrared H-band, where
Nd II atomic lines are sparse and intrinsically weak. 
Atomic architecture also plays a significant role in why Ce acts as a robust control 
group while Nd fails. The Ce II transitions in the H-band benefit from secure experimental
 transition probabilities derived from high-resolution Fourier transform spectrometry 
\citep{lawleretal2009}. Conversely, the extremely weak Nd II lines used by ASPCAP lack similar 
high-precision near-IR laboratory calibrations, forcing the pipeline to rely on theoretical
 line lists with highly uncertain oscillator strengths \citep{denhartogetal2003}. This 
single-window fitting limitation significantly amplifies baseline noise, explaining the 
vertical stretching observed in our cross-match.

The inclusion of Nd in SDSS-V DR19, after its exclusion from DR17, highlights that 
these infrared lines operate at the absolute sensitivity threshold of automated surveys. 
In earlier data releases, Nd lines were often omitted or given zero weight 
because their shallowness threatened to destabilise the global $\chi^2$ 
minimisations for macroscopic parameters. The transition to DR19 introduced 
upgraded atomic data repositories and expanded synthetic grids (using {\sc synspec} and
{\sc turbospectrum}), making it technically possible to map these weaker profiles 
\citep{meszarosetal2025}.
However, the \texttt{Fair} designation in the DR19 science verification papers serves as an 
important boundary condition. While the new linelists enabled measurement, the 
automated global grid optimiser still experiences a dynamic range expansion when processing 
such shallow transitions against high-resolution optical baselines. This could highlight 
a  limitation in translating automated infrared heavy-element spectroscopy to 
classical optical standards.

The systemic scale expansion found in this work has profound implications for our 
understanding of Galactic archaeology. We note that while the literature gradients 
\citep{magrinietal2023, ottoetal2026} are derived using OCs, which occupy a restricted 
parameter space, the present cross-survey sample consists predominantly of field stars covering a
 much broader range of metallicities and kinematics. This allows us to sample the unphysical 
pipeline behavior over a wider dynamic range.
Our findings directly address the contradiction 
between the radial [Nd/H] slopes reported in recent literature. {\it Gaia}-ESO 
\citep{magrinietal2023} reported a global radial [Nd/H] slope of -0.015 $\pm$ 0.003 dex/kpc, 
whereas SDSS-V DR19 \citep{ottoetal2026} reported an [Fe/H] gradient of -0.075 and an 
[Nd/Fe] gradient of -0.052, which leads to a [Nd/H] gradient of -0.127 dex/kpc.
When the  5.540 scaling factor derived above is applied to the SDSS-V slope, the resulting
 corrected value (-0.127/5.540 $\approx$ -0.023 dex/kpc) becomes remarkably consistent with 
the {\it Gaia}-ESO [Nd/H] gradient. Consequently, the corrected SDSS-V [Nd/Fe] gradient 
([Nd/H]$_{cor}$-[Fe/H] = -0.023-(-0.075)= +0.052 dex/kpc) aligns perfectly with the rising 
optical trends.

By rearranging the present ODR model, we provide a calibration tool to convert near-infrared 
abundances into equivalent true optical values:\\

[Nd/H]$_{cor}$ = 0.181 $\cdot$ [Nd/H]$_{SDSS_DR19}$ + 0.101\\

Crucially, we restrict the application of this calibration strictly to the parameters validated 
in this study: $T_{eff} \ge$ 4300 K, log $g>$ 2.5, $\xi <$ 1.8 km/s, and -0.5 dex $\le$ 
[Fe/H] $\le$ +0.3 dex. Applying this global correction outside these boundaries is not recommended.
This calibration allows for the utilization of massive near-infrared datasets while 
maintaining consistency with high-resolution optical benchmarks. By applying this optically 
anchored calibration to the near-infrared datasets, the negative SDSS [Nd/Fe] gradient is 
expected to  vanish, reconciling infrared observations with both theoretical nucleosynthetic 
expectations and optical survey results.

The impact of this spectroscopic scale expansion on our understanding of Galactic 
chemical evolution is best illustrated by the radial abundance gradients. 
Figure~\ref{fig5} provides the evidence for this effect. 
The original SDSS-V DR19 result from the OCCAM survey \citep{ottoetal2026} shows a 
steeply negative [Nd/Fe] gradient of -0.052 $\pm$ 0.008 dex/kpc (red line). 
This result is opposed to  the rising or positive trends 
reported by optical studies such as {\it  Gaia}-ESO \citep{magrinietal2023}, which 
exhibit a positive [Nd/Fe] slope of +0.039 dex/kpc when the [Fe/H] 
and [Nd/H] gradients are combined (blue line).
As shown by the green dashed line, once the 
empirical scaling factor of 5.54 derived in this work is applied to the 
infrared [Nd/H] data, the negative gradient completely vanishes. 
The corrected SDSS-V slope (+0.052 dex/kpc) becomes remarkably consistent 
with the optical high-resolution baseline. This reconciliation demonstrates 
that the previously reported X-shaped expansion is not a physical Galactic 
signature but a byproduct of the sensitivity of the H-band weak-line analysis 
to baseline compression.

\section{Conclusions}

In this study, we have performed a detailed cross-match analysis between the 
optical {\it Gaia}-ESO Survey (DR5) and the near-infrared SDSS-V/MWM (DR19) 
to reconcile the divergent $s$-process abundance scales in the Galactic disc. 
Our main conclusions are summarised as follows:\\

- We established that the underlying stellar atmospheric frameworks 
    of both surveys are remarkably consistent. The comparison of $T_{eff}$, 
    log $g$, and [Fe/H] reveals an excellent 1:1 correlation with minimal 
    scatter, isolating the source of lanthanide discrepancies to the 
    specific line-fitting stages of heavy elements.\\

- Ce serves as a robust control group, exhibiting a stable 
    1:1 relationship between the optical and infrared regimes. This is 
    attributed to the high-contrast Ce II transitions in the H-band 
    and the multi-line window approach employed by the ASPCAP pipeline.\\

- In contrast, Nd exhibits a significant systematic 
    divergence characterized by an X-shaped vertical stretching. The 
    SDSS-V pipeline expands intrinsic Nd variations by a scaling factor 
    of 5.54 relative to the stable {\it Gaia}-ESO optical baseline.\\

- Through synthetic spectral modelling, we identified that this scale 
    expansion is driven by baseline compression. In the near-infrared 
    regime, the growth of the molecular forest (CN and CO) at high 
    metallicities depresses the local continuum. To reach mathematical 
    convergence, the $\chi^2$ minimization engine is forced to artificially 
    inflate the derived abundances of weak features like Nd II.\\

- We have provided an empirical calibration tool, which allows 
  researchers to convert near-infrared abundances 
    into equivalent optical standards, ensuring consistency across surveys.\\

- This correction successfully reconciles the contradictory 
    Galactic radial [Nd/Fe] gradients. Once corrected, the  
    negative gradient reported in SDSS-V results vanishes, aligning 
    with the rising trends established by optical high-resolution 
    benchmarks. This confirms that the previously reported negative slope 
    was a spectroscopic artefact rather than a physical Galactic trend.\\

These results provide a unified roadmap for future high-resolution 
spectroscopic surveys, highlighting the necessity of anchoring automated 
infrared pipelines to established optical baselines when analysing weak 
neutron-capture features.

\begin{acknowledgements}
We thank the referee for the thorough reading of the manuscript and
timely suggestions to improve it.

Data used in this work are available upon request to the author.

\end{acknowledgements}


\begin{appendix}
\onecolumn

\section{Comparison of [Ce/H] and [Nd/H] values in {\it Gaia}-ESO and
SDSS DR19 catalogues.}

\begin{figure}[h!]
\includegraphics[width=4.5cm]{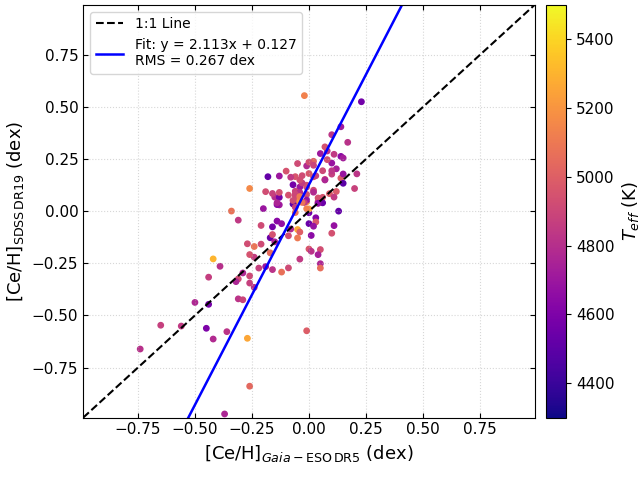}
\includegraphics[width=4.5cm]{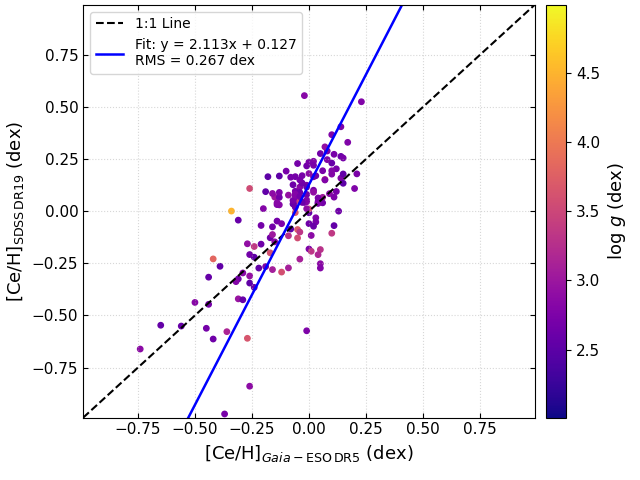}
\includegraphics[width=4.5cm]{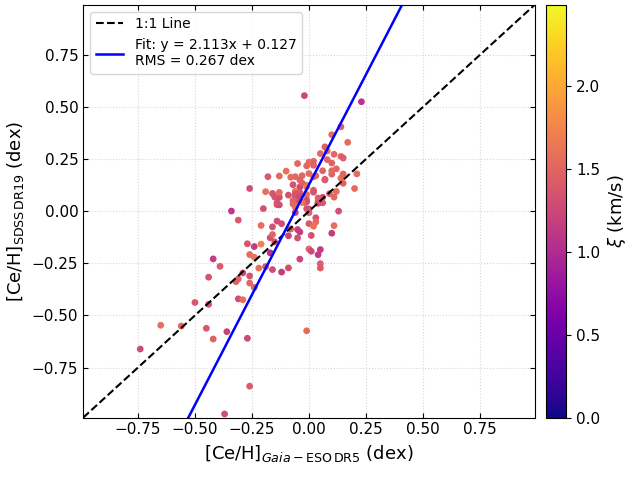}
\includegraphics[width=4.5cm]{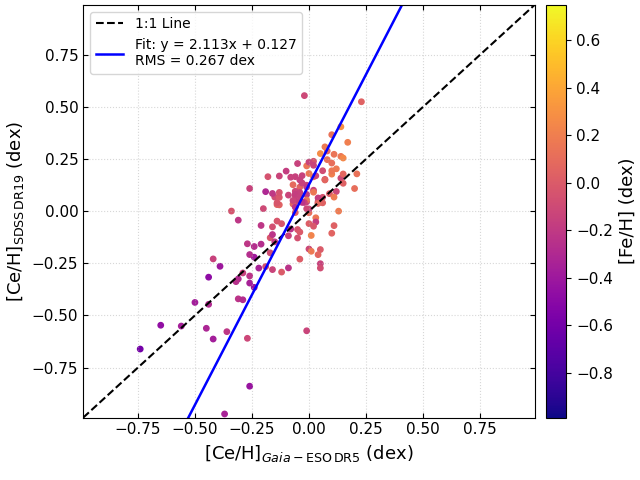}
\caption{Same as Figure~\ref{fig2} and for $T_{eff}$ $<$ 5500 K.}
\label{fig2apend}
\end{figure}

\begin{figure}[h!]
\includegraphics[width=4.5cm]{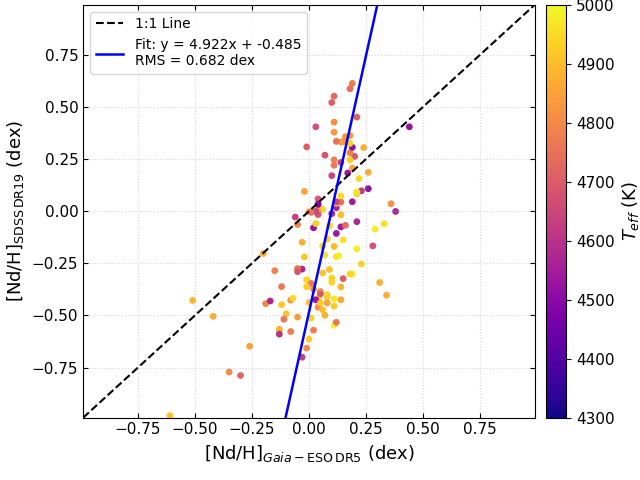}
\includegraphics[width=4.5cm]{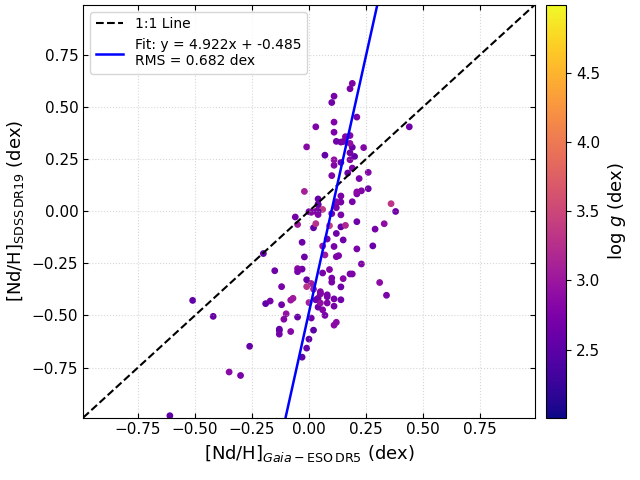}
\includegraphics[width=4.5cm]{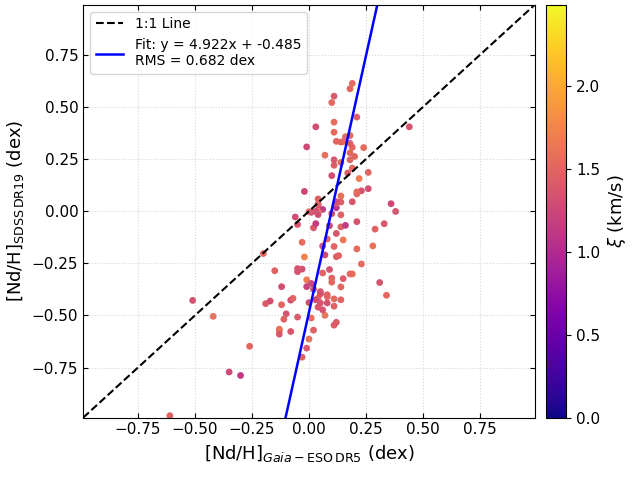}
\includegraphics[width=4.5cm]{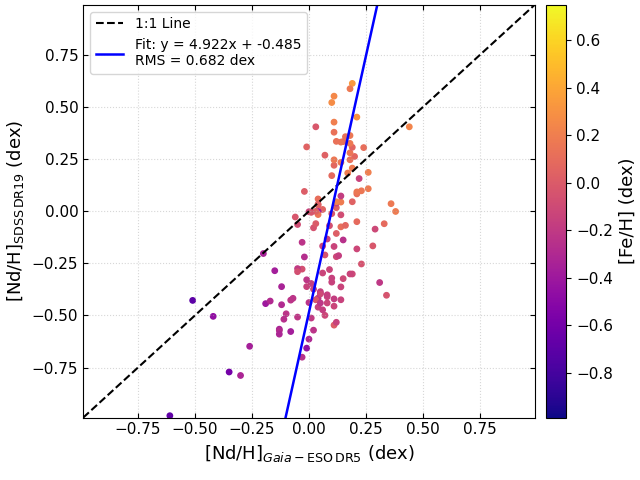}
\caption{Same as Figure~\ref{fig3} and for $T_{eff}$ $<$ 5000 K.}
\label{fig3apend}
\end{figure}

\begin{figure}[h!]
\includegraphics[width=8cm]{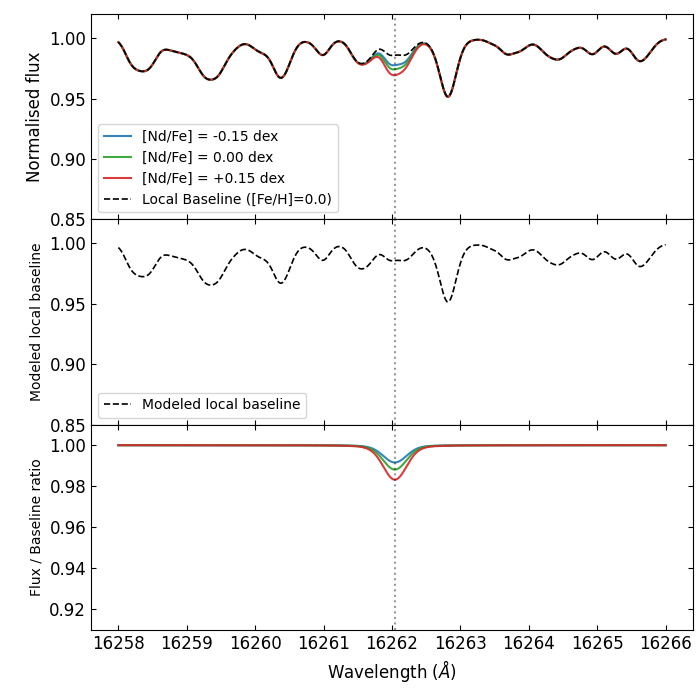}
\caption{Same as Figure~\ref{fig4} for [Fe/H] = 0.0 dex
and different [Nd/Fe] values.}
\label{fig4apend}
\end{figure}

\twocolumn
\end{appendix}

\end{document}